\documentclass{PoS}

\title{Simulation of supersymmetric models on the lattice without a sign problem}
\ShortTitle{Simulating supersymmetric models without a sign problem}

\author{David Baumgartner$^*$ and Urs Wenger$^*$ \\
        Albert Einstein Center for Fundamental Physics\\
        Institute for Theoretical Physics\\
        University of Bern\\
        Sidlerstrasse 5\\
        CH--3012 Bern\\
        Switzerland\\
        E-mail: \email{baumgart@itp.unibe.ch},
        \email{wenger@itp.unibe.ch}}
\author{\phantom{\speaker{David Baumgartner and Urs Wenger}}}

\abstract{ Simulations of supersymmetric models on the lattice with
  (spontaneously) broken supersymmetry suffer from a fermion sign
  problem related to the vanishing of the Witten index. We propose a
  novel approach which solves this problem in low dimensions by
  formulating the path integral on the lattice in terms of fermion
  loops. The formulation is based on the exact hopping expansion of
  the fermionic action and allows the explicit decomposition of the
  partition function into bosonic and fermionic contributions. We
  devise a simulation algorithm which separately samples the fermionic
  and bosonic sectors, as well as the relative probabilities between
  them. The latter then allows a direct calculation of the Witten
  index and the corresponding Goldstino mode. Finally, we present
  results from simulations on the lattice for the spectrum and the
  Witten index for ${\cal N}=2$ supersymmetric quantum mechanics.  }

\FullConference{The XXVIII International Symposium on Lattice Field Theory\\
                 June 14-19,2010\\
                 Villasimius, Sardinia Italy}

\def\Dslash{\slash \hspace{-7pt} D}
\newcommand{\dslash}{\partial \hspace{-6pt}\slash\hspace{+5pt}}

\begin{document}
\section{Introduction}
Supersymmetry (SUSY) is thought to be a crucial ingredient in the
unification of the Standard Model interactions as well as in the
solution of the hierarchy problem. On the other hand, we know that the
low energy physics is not supersymmetric, and consequently the SUSY
must be broken at low energies, either explicitly or
spontaneously. Since the origin and mechanism of spontaneous SUSY
breaking is a non-perturbative effect, it can not be understood in
perturbation theory -- instead, non-perturbative methods are
required. One way to study non-perturbative effects in quantum field
theories is provided by the lattice regularisation. However, the
lattice discretisation comes with various problems, for example the
explicit breaking of Poincar\'e symmetry, or the absence of Leibniz'
rule, and it is therefore not clear at all whether and how SUSY can be
realised within the lattice regularisation.

If the lattice discretisation, for example, enjoys some exact
symmetries that allow only irrelevant symmetry breaking operators,
which become unimportant in the infrared regime, so-called
'accidental' symmetries may emerge from a non-symmetric lattice action
and the full symmetry develops in the continuum limit. This is for
example the case for (Euclidean) Poincar\'e symmetry in lattice QCD or
SUSY in ${\cal N}=1$ SU($N$) Super-Yang-Mills theory. In the latter
case, the only relevant symmetry breaking operator is the gaugino mass
term which violates the $Z_{2N}$ chiral symmetry. Therefore, a
chirally invariant lattice action forbids such a term and SUSY is
automatically recovered in the continuum limit.  For SUSY theories
involving scalar fields, however, such a way out is not available: the
scalar mass term $m^2 |\phi|^2$ breaks SUSY, and there is no other
symmetry available to forbid that term. In principle, even in such
cases, some symmetries can be obtained in the continuum by fine tuning
the theory with counterterms. The restoration of chiral symmetry for
Wilson fermions is one such example. For SUSY, such an approach is in
general not practical, but in lower dimensions, when theories are
superrenormalisable, it sometimes is \cite{Golterman:1988ta}. Yet
another approach to SUSY on the lattice is to look for an exact
lattice realisation of a subalgebra of the full SUSY algebra, e.g.~by
combining the Poincar\'e and flavour symmetry group, so-called twisted
SUSY (cf.~\cite{Catterall:2010jh} and references therein). This
approach is applicable to systems with extended SUSY and leads to
so-called $Q$-exact discretisations.

Yet another difficulty, and maybe the most severe for supersymmetry on
the lattice, is the fact that supersymmetric models with broken
supersymmetry inherently suffer from a fermion sign problem that
hinders Monte Carlo simulations of such models on the lattice. This
can easily be seen as follows. The vanishing of the Witten
index
\[
W \equiv \lim_{\beta \rightarrow \infty}\textrm{Tr} (-1)^F \exp(-\beta H),
\] 
where $F$ is the fermion number and $H$ the Hamiltonian of the system,
provides a necessary (but not sufficient) condition for spontaneous
supersymmetry breaking. On the other hand, the index is
equivalent to the partition function with periodic boundary
conditions,
\[
W = \int_{-\infty}^{\infty} {\cal D}\phi 
\,\, \det\left[\,\Dslash(\phi)\right] \,
e^{-S_B[\phi]} = Z_{p} \, ,
\]
and the only way for the path integral to vanish is through the
fermionic determinant (or Pfaffian) being indefinite, independent of
the fermion discretisation. Indeed, this has been seen in many studies
of supersymmetric models on the lattice that allow spontaneous
supersymmetry breaking, e.g.~SUSY quantum mechanics with a
supersymmetry breaking superpotential\footnote{Note
  that in the context of SUSY quantum mechanics it is misleading to
  speak of spontaneous or dynamical SUSY breaking; it is rather a
  static breaking determined by the form of the superpotential.} 
\cite{Kanamori:2007ye,Kanamori:2010gw,Kanamori:2010ex},
${\cal N}=16$ Yang-Mills quantum
mechanics \cite{Catterall:2007fp,Anagnostopoulos:2007fw,Hanada:2008gy,Hanada:2010rg},
${\cal N}=1$ Wess-Zumino model in 2D \cite{Catterall:2003ae}, or
${\cal N}=(2,2)$ Super-Yang-Mills in 2D \cite{Giedt:2003ve} (see,
however, also \cite{Kanamori:2008bk,Hanada:2010qg}).

Here we propose a novel approach based on
\cite{Wenger:2008tq,Wenger:2009mi} that circumvents the fermionic sign
problem by formulating the path integral on the lattice in terms of
fermion loops. The formulation is based on the exact hopping expansion
of the fermionic action and allows the explicit decomposition of the
partition function into bosonic and fermionic
contributions. Consequently, one can then devise a simulation
algorithm that separately samples the fermionic and bosonic sectors,
as well as the relative probabilities between them. This then allows a
precise calculation of the Witten index and a direct determination of
the presence or absence of a Goldstino mode.  Furthermore, although
this is less relevant in the present context, the approach eliminates
critical slowing down and also allows simulations directly in the
massless limit or at negative bare mass values \cite{Wenger:2008tq}.

\section{Fermion sign problem from spontaneous SUSY breaking}
Let us briefly elaborate further on the issue of spontaneous SUSY
breaking (SSB), the vanishing of the Witten index and the connection
to the fermion sign problem.

It is well known that the Witten index provides a necessary, but not
sufficient, condition for SSB \cite{Witten:1982df}. One has
\[
W \equiv \lim_{\beta \rightarrow \infty}\textrm{Tr} (-1)^F \exp(-\beta H)  \quad
\Rightarrow
\left\{
\begin{array}{lr}
= 0 & \, \textrm{SSB may occur},\\
\neq 0 & \, \textrm{no SSB}.
\end{array}
\right. 
\]
From the definition it is clear that the index counts the difference
between the number of $n_B$ bosonic and $n_F$ fermionic zero energy states:
\[
W \equiv \lim_{\beta \rightarrow \infty} \left[ \textrm{Tr}_B \exp(-\beta  H) 
- \textrm{Tr}_F \exp(-\beta H) 
\phantom{e^H} \hspace{-11pt}
\right] = n_B - n_F \, .
\]
In a field theoretic language the index is equivalent to the partition
function of the system with periodic boundary conditions imposed on
both the bosonic and fermionic degrees of freedom,
\[
W = \int_{-\infty}^{\infty} {\cal D}\phi 
\,\, \det\left[\,\Dslash(\phi)\right] \,
e^{-S_B(\phi)} = Z_{p} \, .
\]
Here the determinant has been obtained by integrating out the complex
valued Dirac fermion fields\footnote{In case one is dealing with
  real-valued Majorana fermion fields, one obtains the Pfaffian
  $\textrm{Pf}\left[\slash \hspace{-5pt} D(\phi)\right]$ instead of the
  determinant.}, while $S_B(\phi)$ is the action for the bosonic
degrees of freedom, collectively denoted by $\phi$. It is now clear
that in order to obtain a vanishing Witten index, we need both
positive and negative contributions to the path integral, and this can
only be achieved by the fermion determinant being indefinite. This is
the source of the fermion sign problem in the context of spontaneous
SUSY breaking, and we argue that such a sign problem must occur in any
model aspiring to accommodate spontaneous SUSY breaking.

It is instructive to illustrate the argument in the explicit example
of SUSY quantum mechanics. The continuum action of $\mathcal{N} = 2$
supersymmetric quantum mechanics reads
\begin{equation}
 S = \int dt  \ \frac{1}{2}\left( \frac{d \phi(t)}{dt}\right)^2 +
 \frac{1}{2}P'(\phi(t))^2 +
 \bar{\psi}(t)\left(\frac{d}{dt}+P''(\phi(t))\right)\psi(t) \, ,
\label{eq:N2SUSYQM}
\end{equation}
where the real field $\phi$ denotes the bosonic coordinate, while
$\bar{\psi}$ and $\psi$ denote the two fermionic coordinates.
$P(\phi)$ is the superpotential and the derivatives $P'$ and $P''$ are
taken with respect to $\phi$. The (regulated) fermion determinant with
periodic boundary conditions can
be calculated analytically \cite{Cooper:1994eh,Bergner:2007pu},
\[
\det\left[\frac{\partial_t + P''(\phi)}{\partial_t +
    m}\right]_{p} = \sinh
  \int_0^T \frac{P''(\phi)}{2} dt \, , 
\]
and by rewriting the $\sinh$-function in terms of two exponentials,
one can separate the positive and negative, or rather, the bosonic and
fermionic contributions to the partition function,
\[
\det\left[\frac{\partial_t + P''(\phi)}{\partial_t + m}\right]_p =
\frac{1}{2} \exp \left(+ \int_0^T \frac{P''(\phi)}{2} dt\right) \, - \, \frac{1}{2}
\exp \left( - \int_0^T \frac{P''(\phi)}{2} dt\right)  \quad
\Longrightarrow  \quad Z_0 - Z_1 \, .
\]
As an example, consider the superpotential $P_e(\phi) = \frac{1}{2}m
\phi^2 + \frac{1}{4} g \phi^4$, which is even under the parity
transformation $\phi \rightarrow \tilde \phi = - \phi$. In this case
one finds $P_e''(\phi) \geq 0$ and hence $Z_0 \neq Z_1$, i.e.~no SUSY
breaking. On the other hand, for the superpotential $P_o(\phi) = -\frac{\mu^2}{4\lambda}
\, \phi + \frac{1}{3} \lambda \phi^3$, which is odd under parity, one
has $P_o''(\tilde \phi) = - P_o''(\phi)$ and, since $S_B(\tilde \phi)
= S_B(\phi)$, one finds $Z_0 = Z_1$, i.e.~a vanishing Witten index and
the corresponding SUSY breaking. Here, the vanishing
of the Witten index is guaranteed by the fact that to each
configuration $\phi$ there exists another configuration $\tilde \phi$
(the parity transformed one) that contributes to the path integral
with the same weight, but with opposite sign stemming from the fermion
determinant. Furthermore, $Z_0 = Z_1$
means that (in the limit of zero temperature, i.e.~$\beta \rightarrow
\infty$) the free energies of the bosonic and fermionic vacua are
equal, and that proves the existence of a massless, fermionic mode
connecting the two vacua, i.e.~the Goldstino mode.

Turning now to SUSY quantum mechanics on the lattice one obtains with a
Wilson type discretisation (cf.~next section for further details)
\begin{equation}
\det\left[\nabla^* + P''(\phi)\right]_{p} = \prod_t \left[1 + P''(\phi_t) \right] - 1 \,,
\label{eq:detp_lattice}
\end{equation}
where $t$ now denotes a discrete lattice site index and $\nabla^*$ is
the backward derivative. Also in this case one can identify the
bosonic and fermionic contributions (i.e.~the first and second term of
the difference in eq.(\ref{eq:detp_lattice})), and we will show below
that this separation is always explicit in the fermion loop
formulation. As a side remark, let us note that in the limit of zero
lattice spacing one finds
\[
\lim_{a \rightarrow 0} \, \det\left[\nabla^* + P''\right]
\quad \longrightarrow \quad \exp \left( +\int_0^T
\frac{P''(\phi)}{2} dt \right) \, \, \,  \det\left[\partial_t +
  P''(\phi)\right] \, ,
\]
where the exponential term can be understood as coming from radiative
contributions that need to be corrected by 'fine-tuning' a
corresponding counterterm \cite{Giedt:2004vb,Bergner:2007pu}.
Reconsidering the two examples for the superpotential mentioned above,
we find for $P_e$ with $m > 0$ and $g\geq 0$ that
\[
\det\left[\nabla^* + P_e''\right] = \prod_t \left[1 + m + 3 g \phi_t^2
\right] - 1 > 0 \, ,
\]
while for $P_o$ one finds 
%
\begin{equation}
\det\left[\nabla^* + P_o''\right] = \prod_t \left[1 + 2 \lambda \phi_t
\right] - 1 \, ,
\label{eq:det_Po}
\end{equation}
which turns out to be indefinite, even when $\lambda > 0$.  While this
is necessary in order to enable a vanishing Witten index, it imposes a
serious problem on any Monte Carlo simulation, for which positive
weights are strictly required. Moreover, the sign problem is severe in
the sense that towards the continuum limit (i.e.~when the lattice
volume goes to infinity), the fluctuations of the first summand in
eq.(\ref{eq:det_Po}) around 1 tend to zero, such that $W \rightarrow
0$ is exactly realised in that limit.  Hence, the source of the
fermionic sign problem lies in the exact cancellation of the first and
the second summand in eq.(\ref{eq:det_Po}), i.e.~of the bosonic and
fermionic contribution to the partition function, and this observation
also holds more generally in higher dimensions.

In the loop formulation, to be discussed in the next section, the
separation of the partition function into the various fermionic and
bosonic sectors is made explicit and allows the construction of a
simulation algorithm that samples these sectors separately, and more
importantly also samples the relative weights between them. In this
way, the loop formulation eventually provides a solution to the
fermion sign problem.

\section{Loop formulation and separation of fermionic and bosonic sectors}
In this section we illustrate the loop formulation and the separation
of the partition function into its fermionic and bosonic sectors by
means of the ${\cal N}=1$ Wess-Zumino model in two dimensions and
$\mathcal{N} = 2$ supersymmetric quantum mechanics in one dimension.

\subsection{${\cal N}=1$ Wess-Zumino model in two dimensions}
The Lagrangian of the  ${\cal N}=1$ Wess-Zumino model in two
dimensions is given by
\begin{equation}
{\cal L} = \frac{1}{2} \left(\partial_\mu \phi \right)^2
+ \frac{1}{2} P'(\phi)^2 
+ \frac{1}{2}\bar{\psi} \left( \dslash + P''(\phi) \right) \psi
\, ,
\label{eq:N1-WZ_Lagrangian}
\end{equation}
where $\psi$ is a real, two-component Majorana field, $\phi$ a real
bosonic field and $P(\phi)$ an arbitrary superpotential. Integrating
out the fermionic Majorana fields yields a Pfaffian which in general,
as discussed above, is not positive definite.

On the lattice, one can use the exact reformulation of the fermionic
Majorana degrees of freedom in terms of non-intersecting,
self-avoiding loops, in order to separate the contributions of the
Pfaffian to the various bosonic and fermionic sectors of the partition
function. A similar exact reformulation of the bosonic degrees of
freedom in terms of bonds, can also be accomplished
\cite{Prokof'ev:2001zz}. While this is not necessary for the solution
of the sign problem, it provides a convenient way to simulate also
those degrees of freedom without critical slowing down, and hence we
will discuss this construction below.

Employing the Wilson lattice discretisation for the fermionic part of the
Lagrangian in eq.(\ref{eq:N1-WZ_Lagrangian}) yields 
\[
 {\cal L}_F = \frac{1}{2} \xi^T {\cal C} (\gamma_\mu \tilde \nabla_\mu
 - \frac{1}{2} \nabla_\mu^* \nabla_\mu + P''(\phi)) \xi \, ,
\]
where $\xi$ now represents the real, 2-component Grassmann field,
while ${\cal C} = -{\cal C}^T$ is the charge conjugation matrix and $
\nabla_\mu^* \nabla_\mu$ the Wilson term.  Using the nilpotency of
Grassmann elements one can expand the Boltzmann factor leading to
\[
\int {\cal D}\xi \, \prod_x \left(1 - \frac{1}{2}M(\phi_x) \xi^T_x
{\cal C}  \xi_x \right) \prod_{x,\mu}\left(1 + \xi^T_x {\cal C}
\Gamma(\mu) \xi_{x+\hat \mu}\right) \, ,
\]
where $M(\phi_x) = 2 + P''(\phi_x)$, $\Gamma(\pm \mu) =
\frac{1}{2}(1\mp\gamma_\mu)$ and $x$ denotes the discrete lattice site
index. Performing now the integration over the fermion field, at each
site $x$ the fields $\xi^T_x {\cal C}$ and $\xi_x$ must be exactly
paired in order to give a contribution to the path integral, so one
finds
\[
\int {\cal D}\xi \, \prod_x \left(-M(\phi_x) \xi^T_x
{\cal C}  \xi_x \right)^{m(x)} \prod_{x,\mu}\left(\xi^T_x {\cal C}
\Gamma(\mu) \xi_{x+\hat \mu}\right)^{b_\mu(x)} \, ,
\]
where the occupation numbers $m(x)=0,1$ for the monomer terms and
$b_\mu(x)=0,1$ for the fermionic bonds (or dimers), satisfy the
constraint
\begin{equation}
m(x) + \frac{1}{2}\sum_{\mu} \left( b_{\mu}(x) +
  b_{\mu}(x-\hat \mu) \right)= 1 \quad \forall x \, .
\label{eq:D2_constraint}
\end{equation}
This constraint is equivalent to the fact that only closed,
self-avoiding paths survive the Grassmann integration.  When
integrating out the fermion fields, the projectors $\Gamma(\mu)$
eventually yield a weight $\omega$, which only depends on the
geometric structure of the specific constrained path (CP)
configuration $\ell \in {\cal L}$. In particular, one has
\[
|\omega(\ell)| = 2^{-n_c/2} \, ,
\]
where $n_c$ denotes the number of corners in the loop configuration,
while the sign depends on the topology of the loop configuration and
will be discussed below.

As mentioned above, the bosonic fields can be treated analogously
\cite{Prokof'ev:2001zz}. On the lattice, the kinetic term
$(\partial_\mu \phi)^2$ yields $\phi_x \phi_{x-\hat\mu}$, and
expanding this hopping term to all orders gives
\begin{equation}
\int {\cal D}\phi \prod_{x,\mu} \sum_{n_\mu(x)} \frac{1}{n_\mu(x)!}  \left(\phi_x
  \phi_{x-\hat\mu}\right)^{n_\mu(x)} \, \prod_{x} \exp\left( -\frac{1}{2} V(\phi_x) \right) \, M(\phi_x)^{m(x)}
\label{eq:bosonic_hopping_expansion}
\end{equation}
with bosonic bond occupation numbers $n_\mu(x) = 0,1,2,\ldots$ and
$V(\phi_x) = 4 + P'(\phi_x)^2$. In contrast to the fermionic case, the
exact reformulation requires one to include an infinite number of
terms in the hopping expansion, and hence occupation numbers up to
infinity, instead of just 0 and 1 as for the fermionic
bonds. Integrating out the bosonic fields $\phi_x$ yields the site
weights
\[
Q(n_\mu(x),m(x)) = \int d\phi_x \, \,\exp\left( - \frac{1}{2} V(\phi_x) \right) \,  \phi_x^{N(x)}  M(\phi_x)^{m(x)},
\]
where $N(x) = \sum_\mu \left(n_\mu(x) + n_\mu(x-\hat \mu) \right)$
counts the number of bosonic bonds attached to a given site, while
$M(\phi_x)^{m(x)}$ may contribute additional powers of $\phi_x$. So
the bosonic contribution to the weight of a given configuration
factorises into a product of local weights,
\[
W(n_\mu(x),m(x)) = \prod_{x,\mu} \frac{1}{n_\mu(x)!}  \prod_x
Q(n_\mu(x),m(x)) \, .
\]

In summary, the fermionic and bosonic degrees of freedom in our
original partition function are now expressed in terms of fermionic
monomers and dimers and bosonic bonds, and the integration over the
fields has been replaced by a constrained sum over all allowed
monomer-dimer-bond configurations, yielding
\[
Z = 
\sum_{\{\ell\} \in {\cal L}} \sum_{\{n_\mu\} \in CP}
|W(n_\mu(x),m(x)) \cdot \omega(\ell)| \, .
\]

In particular, the partition function for the fermionic degrees of
freedom is represented by a sum over all non-oriented, self-avoiding
fermion loops
\begin{equation}
Z_{\cal L} =  \sum_{\{\ell\}\in{\cal L}} \large| W(n_\mu(x),m(x))
\cdot \omega(\ell) \large|,\quad
{\cal L} \in {\cal L}_{00} \cup {\cal L}_{10} \cup {\cal L}_{01} \cup
{\cal L}_{11} \, ,
\label{eq:ZL}
\end{equation}
where $\ell$ represents a fermion loop configuration in one of the
four topological classes ${\cal L}_{l_1,l_2}$, with $l_1,l_2=0,1$
denoting the total number of loop windings (modulo 2) along the first
and second direction, respectively, on the periodic lattice
torus. $\omega$ denotes the weight of the specific loop configuration
and depends on the geometry of the loop configuration, as discussed
above. The sign of the weight is solely determined by the topological
class and the fermionic boundary conditions $\epsilon_\mu$, where
$\epsilon_\mu = 0,1$ stands for periodic and anti-periodic boundary
conditions, respectively.  Configurations in ${\cal L}_{00}$ have
positive weights independent of the boundary conditions, while the
sign of the weights for configurations in ${\cal L}_{01}, {\cal
  L}_{10}$ and ${\cal L}_{11}$ is given by $(-1)^{l_\mu \cdot
  \epsilon_\mu + 1}$. So the partition function $Z_{\cal L}$ in
eq.(\ref{eq:ZL}), where all sectors contribute positively, represents
a system with unspecified fermionic boundary conditions
\cite{Wolff:2007ip}, while a partition function with fermionic
b.c.~periodic in the spatial and anti-periodic in the temporal
direction, respectively, is described by the combination
\[
Z_{pa} = Z_{{\cal L}_{00}} - Z_{{\cal L}_{10}} + Z_{{\cal L}_{01}} +
Z_{{\cal L}_{11}} \, .
\]
This combination represents the system at finite
temperature. Analogously, the partition function with fermionic
b.c.~periodic in all directions -- the Witten index -- is given by
\[
Z_{pp} = Z_{{\cal L}_{00}} - Z_{{\cal L}_{10}} - Z_{{\cal L}_{01}} -
Z_{{\cal L}_{11}} \, .
\]
The interpretation of the Witten index in terms of the partition
functions $Z_{{\cal L}_{ij}}$ is straightforward. Any fermion loop
winding non-trivially around the lattice carries fermion number $F=1$,
hence configurations with an odd number of windings,
i.e.~configurations in $Z_{{\cal L}_{10}}, Z_{{\cal L}_{01}}$ and
$Z_{{\cal L}_{11}}$, also carry fermion number $F=1$, while
configurations with no, or an even number of windings, i.e.~in
$Z_{{\cal L}_{00}}$, have $F=0$. The partition function $Z_{{\cal
    L}_{00}}$ may therefore be interpreted as representing the bosonic
vacuum, while the combination $Z_{{\cal L}_{10}} + Z_{{\cal L}_{01}} +
Z_{{\cal L}_{11}}$ corresponds to the fermionic vacuum. Consequently,
the latter contributes to the Witten index $W \equiv Z_{pp}$ with
opposite sign relative to the bosonic vacuum. Since each of the four
partition functions is positive, vanishing of the Witten index implies
$Z_{{\cal L}_{00}} = Z_{{\cal L}_{10}} + Z_{{\cal L}_{01}} + Z_{{\cal
    L}_{11}}$.

\subsection{$\mathcal{N} = 2$
supersymmetric quantum mechanics}
The loop formulation for the $\mathcal{N} = 2$ supersymmetric quantum
mechanics on the lattice is obtained in a completely analogous
manner. Using again the Wilson lattice discretisation for the
fermionic part, the continuum action in eq.(\ref{eq:N2SUSYQM}) reads
\[
 S_L  =  \sum_x \frac{1}{2}(P'(\phi_x)^2 + 2\phi_x^2) -
 \phi_x\phi_{x-1} + (1 + P''(\phi_x))\bar{\psi}_x\psi_x -
 \bar{\psi}_x\psi_{x-1} \, ,
\]
where $x$ now denotes the one-dimensional, discrete lattice site
index. Note that in one dimension the fermionic lattice derivative,
including the contribution from the Wilson term with Wilson parameter
$r=1$, becomes a simple, directed hop $\bar{\psi}_x\psi_{x-1}$ which,
in the loop formulation, can be described by the (directed) bond
occupation number $b(x)=0,1$. Integrating out the fermionic degrees of
freedom yields a constraint for the fermion monomer and bond
occupation numbers, analogous to eq.(\ref{eq:D2_constraint}), namely
\[
m(x) + \frac{1}{2}\left(b(x) + b(x-1)\right) = 1 \quad \forall x.
\]
So for $\mathcal{N} = 2$ supersymmetric quantum mechanics the loop
formulation becomes particularly simple: there are just two different
loop configurations, namely a bosonic one, where ${m(x)=1, b(x) = 0}$
for all $x$, and a fermionic one where ${m(x)=0, b(x) = 1}$ for all
$x$. Since the latter corresponds to a closed fermion loop, it will
pick up a minus sign from the Grassmann integration, relative to the
bosonic contribution.

On top of the two fermion loop configurations one may treat the
bosonic fields in the same way as before and employ a hopping
expansion to all orders. After rearranging the bosonic fields the
integration can eventually be performed separately at each site and
one ends up with the weight
\begin{equation}
 W(n(x),m(x))  =  \prod_x \frac{1}{n(x)!} \int d \phi_x \ \phi_x^{n(x) +
   n(x-1)} \ \mathrm{e}^{-\frac{1}{2}V(\phi_x)} (1 + P''(\phi_x))^{m(x)} 
\label{eq:weight_SSQM}
\end{equation}
for a given bosonic and fermionic bond configuration, with $V(\phi_x)
= 2 + P'(\phi)^2$. In terms of these weights the partition function
can now be written as
\begin{equation}                                              
 Z_{\cal L} = \int \mathcal{D} \phi \mathcal{D} \bar{\psi}\mathcal{D} \psi \
 \mathrm{e}^{-S_L} =  \sum_{\{\ell\}\in{\cal L}}              
 \sum_{\{n\}\in \textrm{\tiny CP}}\large| W(n(x),m(x)) \large|,\quad
{\cal L} \in {\cal L}_{0} \cup {\cal L}_{1} \, ,     
\label{eq:ZL_SUSYQM}         
\end{equation}
where the second sum is over all allowed bosonic bond configurations
$\{n\}\in \textrm{CP}$, and $\ell$ represents one of the two fermion
loop configurations in the topological classes ${\cal L}_0$ or ${\cal
  L}_1$, respectively. As before, the sign of the weight is solely
determined by the topological class and the fermionic boundary
condition. If $l=0,1$ denotes the fermion loop winding number and, as
before, $\epsilon=0,1$ the periodic and anti-periodic fermionic
boundary condition, respectively, the sign is given by
$(-1)^{l\cdot(\epsilon+1)}$.

Choosing anti-periodic fermionic boundary conditions $\epsilon=1$ we find 
\[
Z_a = Z_{{\cal L}_0} +  Z_{{\cal L}_1} \, ,
\]
which is simply the partition function for the system at finite
temperature, while choosing periodic fermionic boundary conditions
$\epsilon=0$ yields
\[
Z_p = Z_{{\cal L}_0} -  Z_{{\cal L}_1}
\]
representing the Witten index. Here, the interpretation is
particularly intuitive: the two fermion loop configurations simply
represent the bosonic and fermionic vacuum, while $Z_{{\cal L}_0}$ and
$Z_{{\cal L}_1}$ represent the bosonic and fermionic partition
function in the corresponding sectors. The Witten index vanishes
whenever $Z_{{\cal L}_0} = Z_{{\cal L}_1}$, i.e.~when the
contributions from the bosonic and fermionic sectors cancel. In this
case, the free energy of the bosonic and fermionic vacuum is equal,
and this is equivalent to saying that there exists a gapless,
fermionic excitation which oscillates between the two vacua, i.e.~the
Goldstino mode.

\section{Solution of the fermion sign problem}
In this section we briefly describe the simulation algorithm and
explain how it eventually solves the fermion sign problem.
The loop system can most efficiently be simulated by enlarging the
configuration space by open strings. Following \cite{Prokof'ev:2001zz}
the bosonic bonds are updated by inserting two bosonic sources, which
sample directly the bosonic 2-point correlation function. Similarly,
the fermion bonds are most efficiently updated by simulating a
fermionic string \cite{Wenger:2008tq} that samples the configuration
space of the fermionic 2-point correlation function, instead of the
standard configuration space of loops. 

For the ${\cal N}=1$ Wess-Zumino model in two dimensions, where the
fermionic degrees of freedom are Majorana, the open fermionic string
is non-oriented and corresponds to the insertion of a Majorana fermion
pair $\{\xi^T_x {\cal C},\xi_y\}$ at position $x$ and $y$, while for
${\cal N}=2$ supersymmetric quantum mechanics, where the fermionic
degrees of freedom are Dirac, the string is oriented and corresponds
to the insertion of a Dirac fermion pair $\{\bar \psi_x, \psi_y\}$.

The algorithm proceeds by locally updating the endpoints of the open
fermionic string using a simple Metropolis or heat bath step according
to the weights of the corresponding 2-point function
(cf.~\cite{Wenger:2008tq} for details). When one end is shifted from,
say, $x$ to one of its neighbouring sites $y$, a fermionic dimer on
the corresponding bond is destroyed or created depending on whether
the bond is occupied or not. Contact with the partition functions
$Z_{{\cal L}_{i j}}$ and $Z_{{\cal L}_{i}}$, respectively, is made
each time the open string closes. This then provides the proper
normalisation for the expectation value of the 2-point function.

The solution of the fermion sign problem discussed above relies on the
correct determination of the relative weights between the bosonic and
fermionic sectors, and the simulation algorithm described in
\cite{Wenger:2008tq} achieves this in a most efficient way.  The open
fermionic string tunnels between loop configurations in the various
topological homotopy classes ${\cal L}_{00}, {\cal L}_{10}, {\cal
  L}_{01}, {\cal L}_{11}$ in two, and ${\cal L}_{0}, {\cal L}_{1}$ in
one dimension, thereby determining the relative weights between the
partition functions $Z_{{\cal L}_{00}}, Z_{{\cal L}_{10}}, Z_{{\cal
    L}_{01}}, Z_{{\cal L}_{11}}$, or $Z_{{\cal L}_{0}}, Z_{{\cal
    L}_{1}}$, respectively. From the relative weights, the Witten
index (or any other partition function of interest) can be
reconstructed a posteriori.

Let us emphasise that the open string algorithm and the corresponding
absence of critical slowing down at the critical point as reported in
\cite{Wenger:2008tq} is crucial for the solution of the sign
problem. Since the algorithm updates the configurations according to
the fermionic 2-point correlation function, they are updated equally
well on all length scales up to a scale of the order of the largest
fermionic correlation length. This is in fact the reason why critical
slowing down is essentially absent even at a critical point when the
correlation length becomes infinite.  Now, in order to have $W=0$ in
the continuum ($L \rightarrow \infty$) the Goldstino mode has to
become massless. Since the algorithm ensures an efficient update of
that mode, the tunneling between the bosonic and fermionic vacua is
guaranteed and the Witten index indeed vanishes in practice.

\section{Results  for ${\cal N}=2$
supersymmetric quantum mechanics}
We are now ready to present some results for the spectrum and the
Witten index for the case of ${\cal N}=2$ supersymmetric quantum
mechanics.

\subsection{Spectrum for the perturbatively improved standard discretisation}
Here we present our results for the spectrum using the standard
discretisation including a counterterm. As discussed in
\cite{Giedt:2004vb}, for the standard discretisation described above,
the correct continuum limit is spoiled by radiative
corrections\footnote{Note, however, that in one dimension these
  corrections are finite.}. This can be corrected by adding a
counter\-term of the form $\frac{1}{2}\sum_x P'(\phi_x)$
\cite{Giedt:2004vb}. By doing so, one ensures that all observables
reach the correct continuum limit and that the full supersymmetry is
eventually restored.

\begin{figure}[t]
\includegraphics[width=0.5\textwidth]{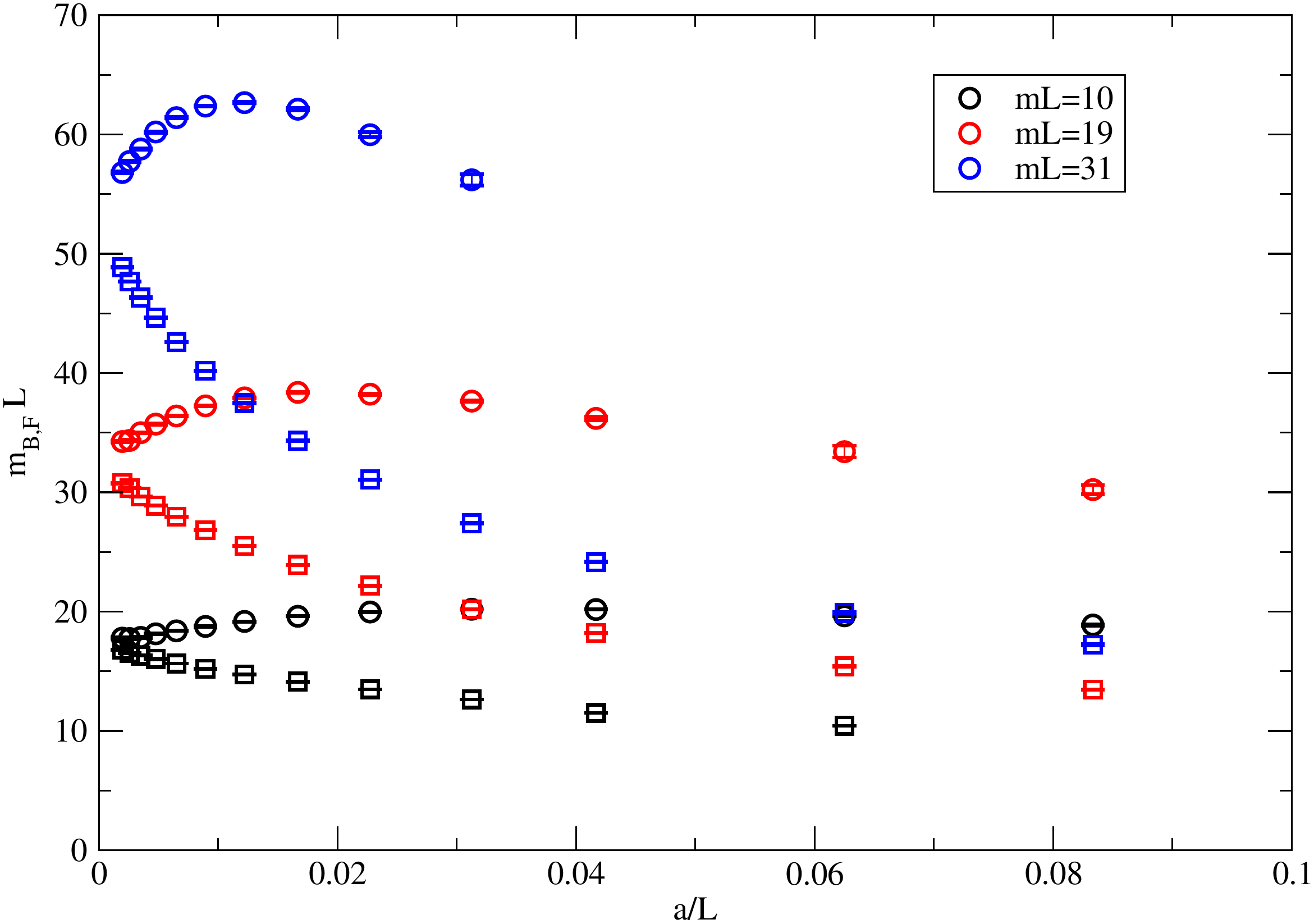}
\includegraphics[width=0.5\textwidth]{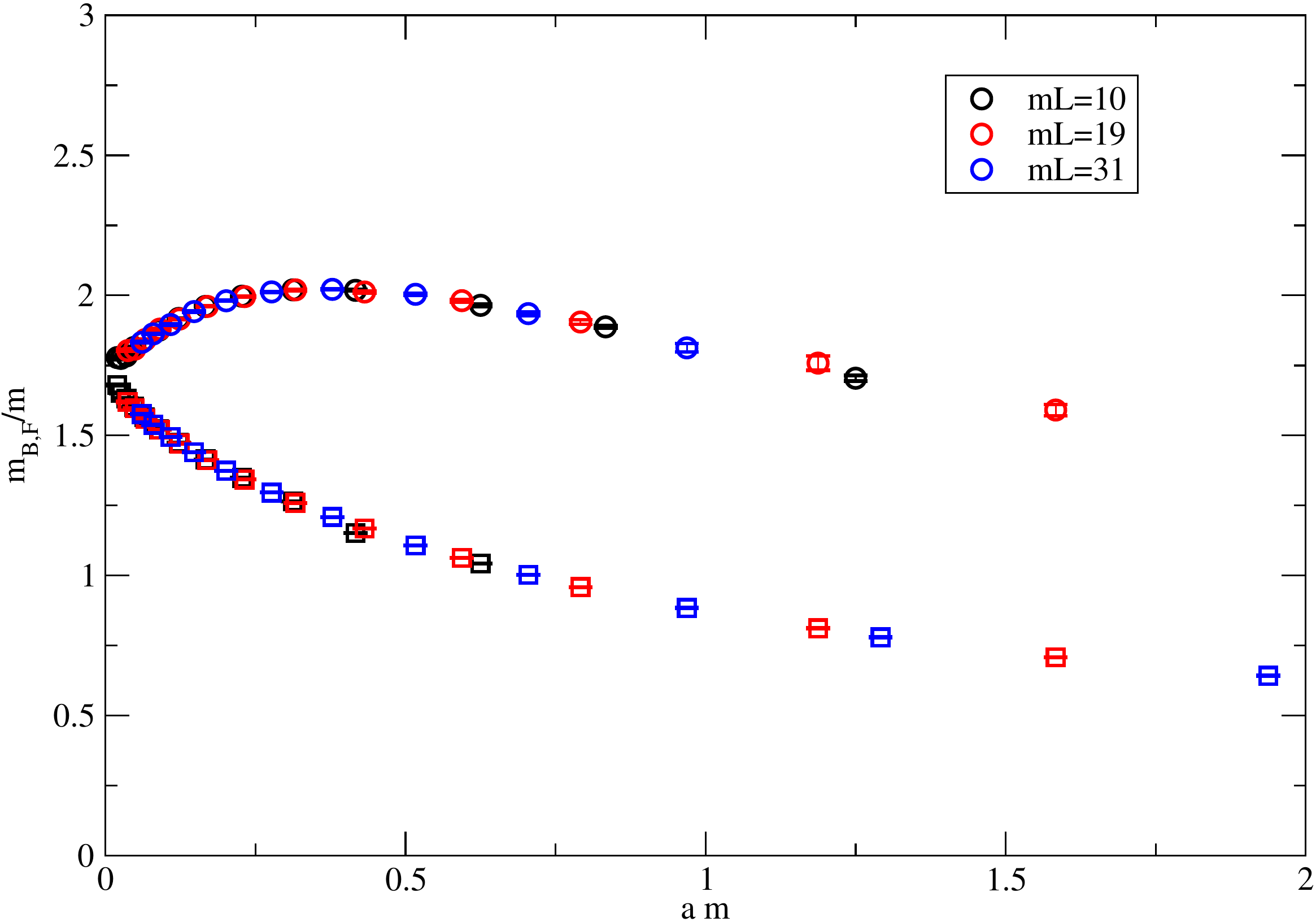}
\caption{Spectrum for the perturbatively improved standard
  discretisation at $g/m^2 = 1.0$. In the left plot the excitation
  energies $m_B$ (circles) and $m_F$ (squares) and the lattice spacing are expressed in
  units of the lattice extent $L$, while in the right plot they are
  expressed in units of the bare mass parameter $m$.}
\label{fig:cl_gm2_1_impr}
\end{figure}

As an example we consider a superpotential with unbroken
supersymmetry, i.e.~$P_e(\phi) = {\frac{1}{2} m \phi^2 + \frac{1}{4}g
  \phi^4}$, and choose the coupling $g/m^2 = 1.0$. The results are
presented in figure \ref{fig:cl_gm2_1_impr} where we show the lowest
lying excitation energies for the boson (circles) and the fermion
(squares) as a function of the lattice spacing $a$ for various values
of fixed $mL$. In the left plot, the quantities are expressed in units
of the lattice extent $L$, while in the right plot, they are expressed
in units of the bare mass parameter $m$ in order to illustrate the
common scaling behaviour.
The leading lattice artifacts turn out to be ${\cal O}(a)$ for both
the fermion and boson masses.  At finite lattice spacing the
supersymmetry is explicitly broken by the discretisation, and hence
the boson and fermion masses are not degenerate. In the continuum
limit, however, the supersymmetry is restored and the masses become
degenerate.

\subsection{Spectrum for the $Q$-exact discretisation}
As briefly discussed in the introduction, for models with extended
supersymmetry it is sometimes possible to preserve some of the
supersymmetries exactly at finite lattice spacing
\cite{Catterall:2000rv}. The so-called $Q$-exact discretisations
preserve a suitable sub-algebra of the full supersymmetry algebra,
i.e.~a linear combination of the available supersymmetries. In the
context of $\mathcal{N} = 2$ supersymmetric quantum mechanics the
$Q$-exact action is obtained from the standard action by adding, e.g.,
the term $\sum_x P'(\phi_x) \nabla^*\phi_x$, but other forms are also
possible \cite{Catterall:2003xx,Bergner:2007pu}.  Since the term
contains a derivative, there are additional hopping terms that need to
be considered in the hopping expansion \cite{baumgartner:2010}.

\begin{figure}[t]
\includegraphics[width=0.5\textwidth]{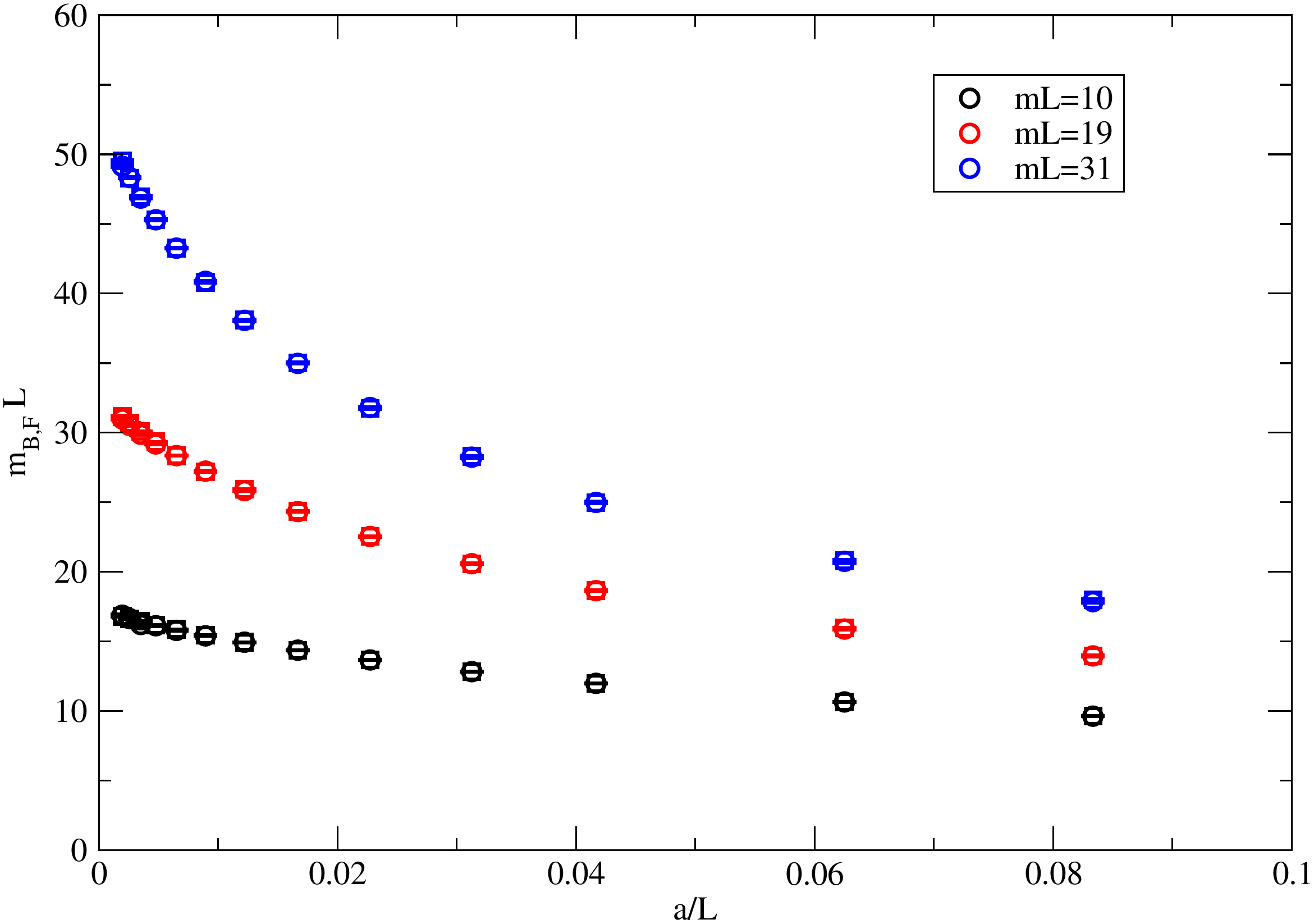}
\includegraphics[width=0.5\textwidth]{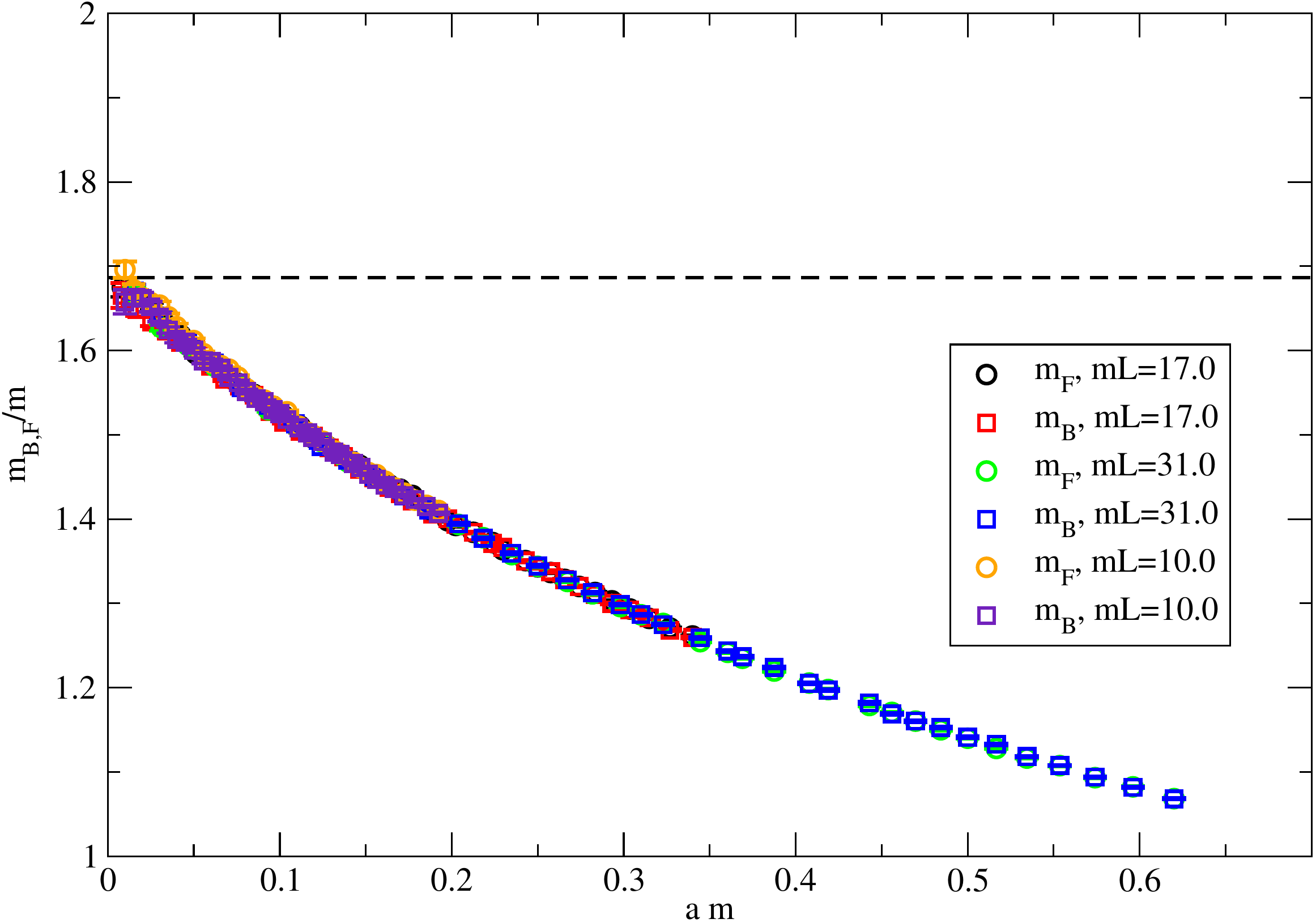}
\caption{Spectrum for the $Q$-exact
  discretisation at $g/m^2 = 1.0$. In the left plot the excitation
  energies $m_B$ (circles) and $m_F$ (squares) and the lattice spacing are expressed in
  units of the lattice extent $L$, while in the right plot they are
  expressed in units of the bare mass parameter $m$. The dashed line
  denotes the exact value obtained with Numerov's method.}
\label{fig:cl_gm2_1_Qexact}
\end{figure}

In the following we concentrate again on the superpotential $P_e$ with
unbroken supersymmetry. Using the $Q$-exact discretisation one expects
degenerate fermion and boson masses even at finite lattice spacing
\cite{Catterall:2000rv} and this is beautifully confirmed by our
results at $g/m^2=1.0$ presented in figure \ref{fig:cl_gm2_1_Qexact}.
Note that the leading lattice artifacts are again ${\cal O}(a)$ for
both the fermion and boson masses.

Finally, in the right plot we show the results of a high precision
simulation that serves the purpose of checking the correctness of the
new simulation algorithm, as well as our procedures for the extraction
of the fermion and boson masses. Indeed, we can confirm the mass
degeneracy to a precision better than a few per mill at all lattice
spacings, and the continuum value of the mass gap agrees with the
exact result in the continuum obtained with Numerov's method (dashed
line) also within a few per mill. Due to the loop formulation and the
efficiency of the new simulation algorithm, these results can be
obtained with a very modest computational effort.

\subsection{Witten index}
\begin{figure}[t]
\includegraphics[width=0.5\textwidth]{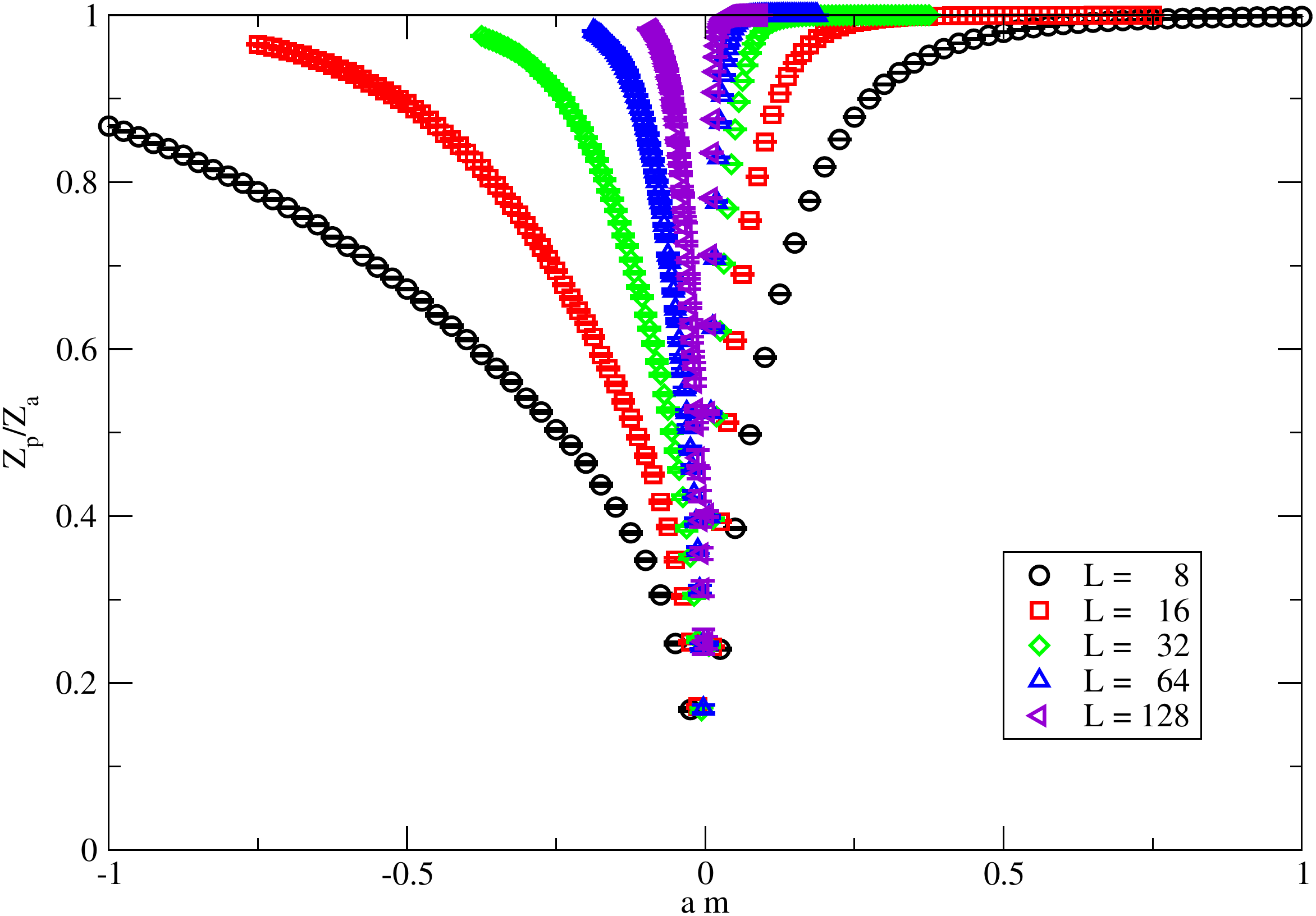}
\includegraphics[width=0.5\textwidth]{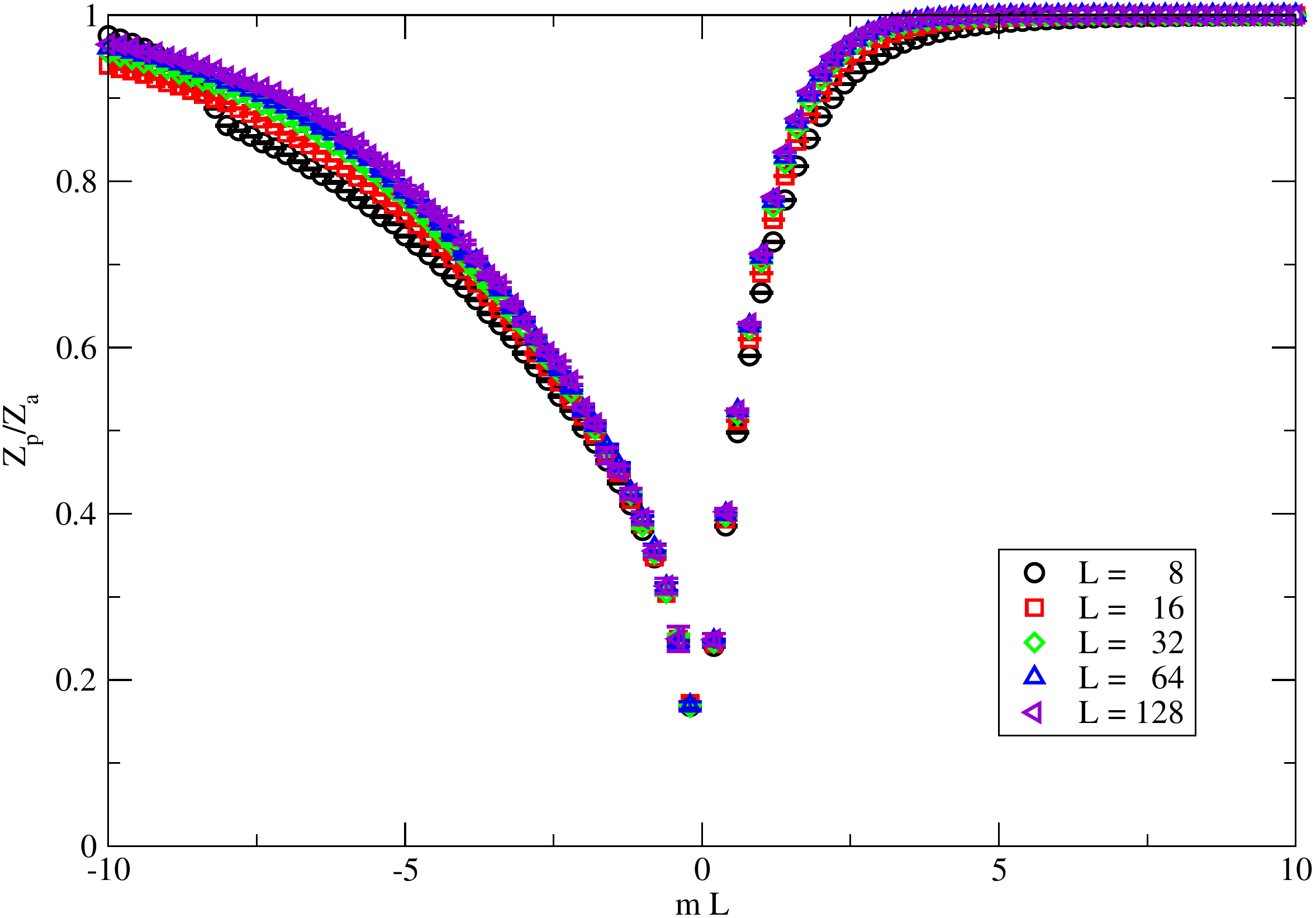}
\caption{Simulation results for the Witten index $W \propto Z_p/Z_a$
  as a function of the bare mass parameter $a m$ (left plot), and as a
  function of the inverse temperature $mL$ (right plot), for a system
  with {\it unbroken} supersymmetry using the $Q$-exact discretisation at
  $g/m^2 = 1.0$. 
}
\label{fig:zratio_Gm2_1_Qexact}
\end{figure}

Let us turn to the Witten index $W \propto Z_p/Z_a$, i.e.~the
partition function with periodic boundary conditions $Z_p$, normalised
to the finite temperature partition function $Z_a$. We start with the
superpotential $P_e(\phi) = \frac{1}{2}m \phi^2 + \frac{1}{4} g
\phi^4$ for which supersymmetry is unbroken and $W \neq 0$. The
results are presented in figure \ref{fig:zratio_Gm2_1_Qexact}, where
we show $Z_p/Z_a$ in the left plot as a function of the bare mass
parameter $a m$ for the coupling $g/m^2 = 1.0$ using the $Q$-exact
discretisation. The continuum limit is reached as $a m \rightarrow 0$,
and we indeed find that $Z_p/Z_a \rightarrow 1$, i.e.~$W\neq0$ in that
limit on a sufficiently large lattice. The fact that at fixed lattice
extent $L$ the partition function ratio goes to zero with $a m
\rightarrow 0$ can be interpreted as a 'finite size', or rather finite
temperature effect, since the temperature $T$ is inversely
proportional to the extent of the lattice $L$. So if we plot the data
as a function of $m/T = m L$, $m L = 0$ corresponds to infinitely high
temperature, while the limit $m L \rightarrow \infty$ corresponds to
$T \rightarrow 0$. This is illustrated in the right plot of figure
\ref{fig:zratio_Gm2_1_Qexact}, from where we find $Z_p/Z_a \rightarrow
1$ in the continuum limit for temperatures $m L \gtrsim 5$.  The data
shows a rather good scaling behaviour towards the continuum limit ($L
\rightarrow \infty$ at fixed $mL$). In the continuum, the value
$Z_p/Z_a = 1$ is approached exponentially fast with $mL$, i.e.~towards
zero temperature $mL \rightarrow \infty$.

Note that with the formulation and algorithm presented here, it is
also possible to simulate at negative bare mass $m<0$. In this case,
we still find $Z_p/Z_a \rightarrow 1$ in the continuum limit towards
zero temperature, albeit at a slower rate.

\begin{figure}[t]
\includegraphics[width=0.5\textwidth]{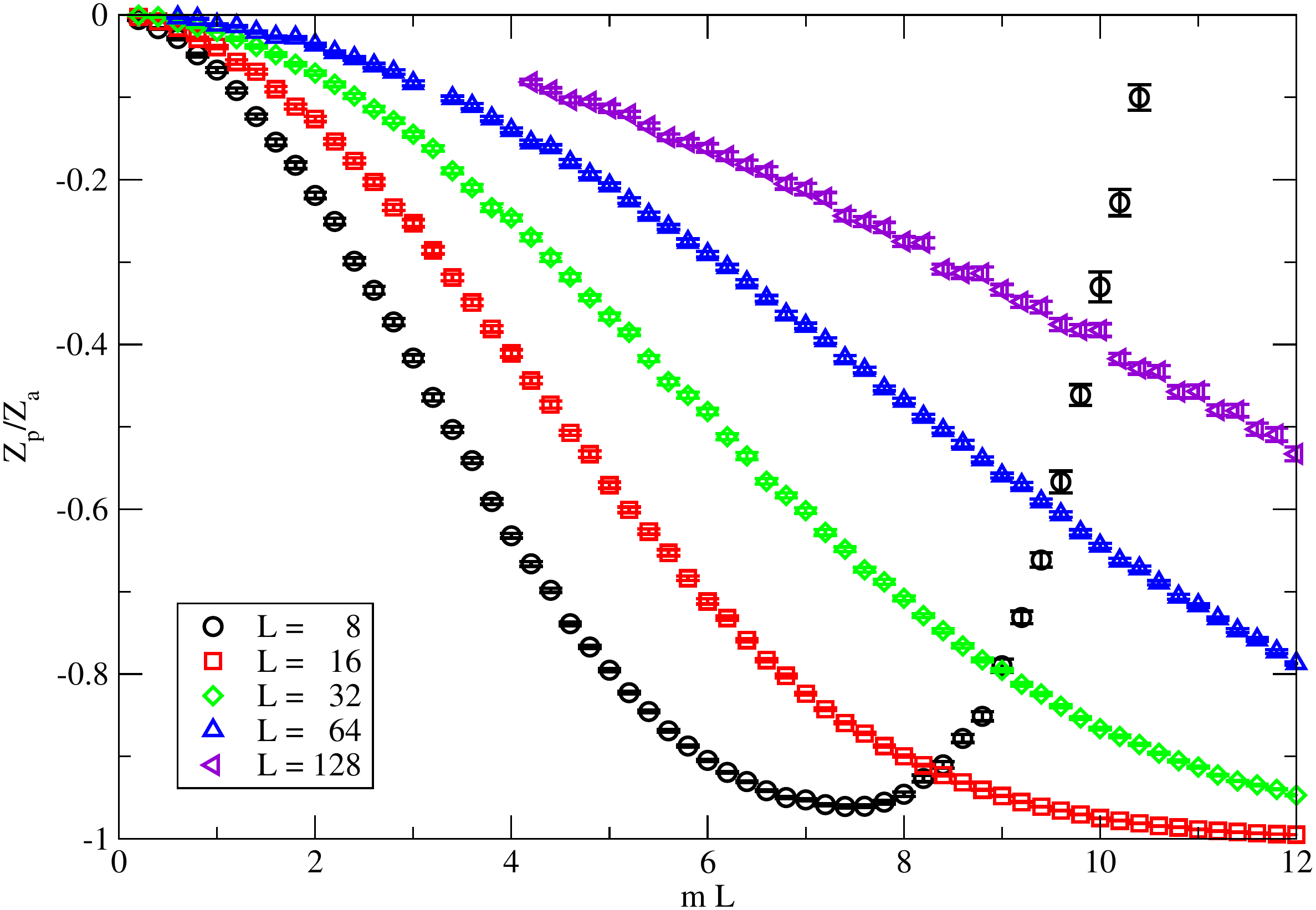}
\includegraphics[width=0.5\textwidth]{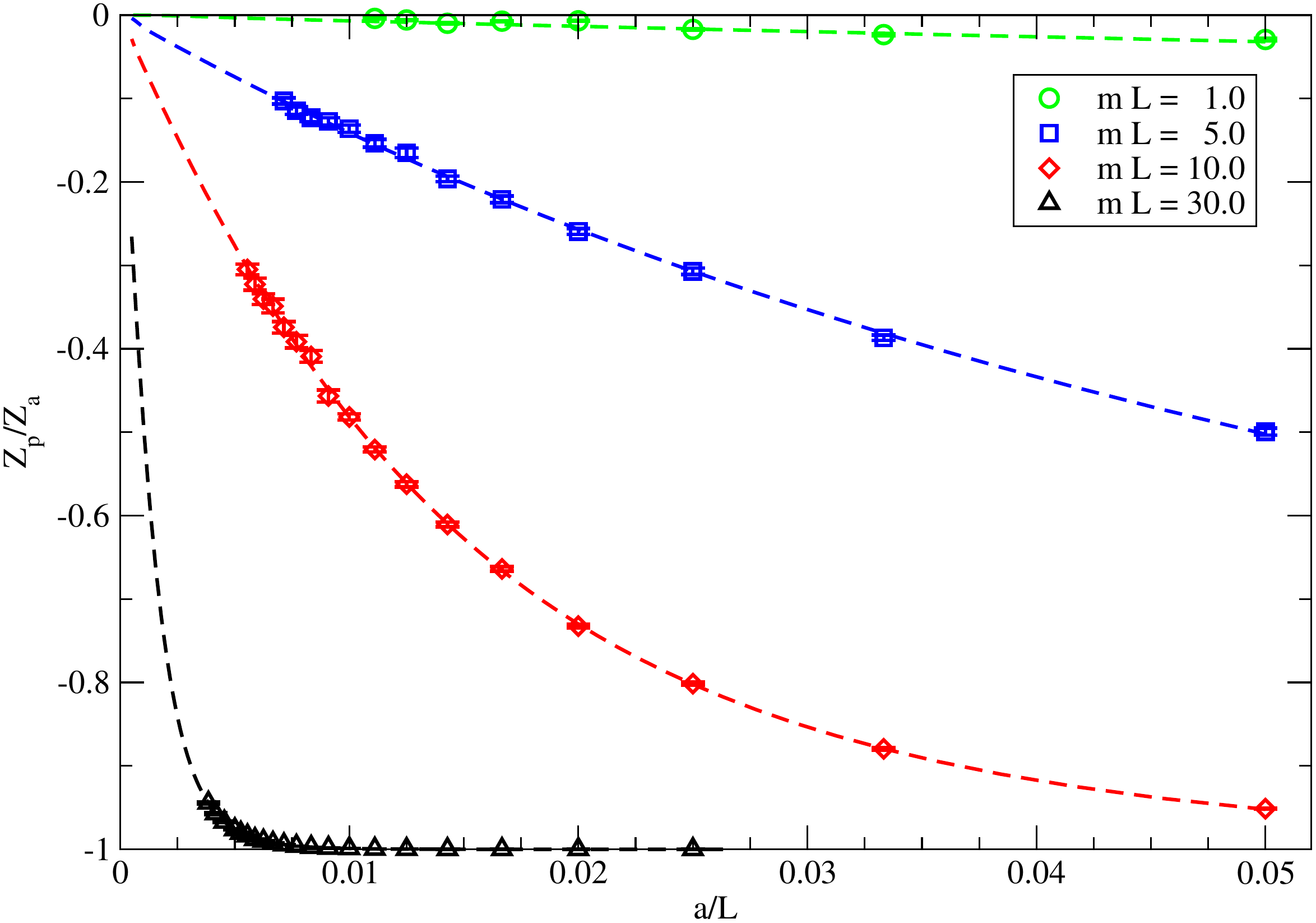}
\caption{Simulation results for the Witten index $W \propto Z_p/Z_a$
  as a function of the inverse temperature $mL$ (left plot), and as a
  function of the lattice spacing $a/L$ at fixed temperature (right
  plot), for a system with {\it broken} supersymmetry using the standard,
  perturbatively improved discretisation at $\lambda^2/m^{3} = 1.0$.}
\label{fig:zratio_Lambdam32_1_impr}
\end{figure}

When repeating this exercise for the superpotential $P_o(\phi) =
-\frac{m^2}{4\lambda} \, \phi + \frac{1}{3} \lambda \phi^3$, for which
supersymmetry is broken, we should expect a vanishing Witten
index. Our results for this case are presented in figure
\ref{fig:zratio_Lambdam32_1_impr}, where we show the Witten index $W
\propto Z_p/Z_a$ as a function of $mL$ in the left plot, and as a
function of the lattice spacing $a/L$ at fixed values of $mL$ in the
right plot using the standard, perturbatively improved discretisation
at $\lambda^2/m^{3} = 1.0$.
In this case we find that while the Witten index $W \propto Z_p/Z_a$
approaches zero at 'infinite temperature' $mL \rightarrow 0$ as in the
unbroken case, it also does so at any value of $mL$ in the continuum
limit, even though for large values of $mL$ the scaling towards the
continuum limit is reached only at very fine lattice spacings. In
fact, for large $mL$ the approach $W \rightarrow 0$ is exponentially
slow in the lattice spacing. This is exemplified in the right plot of
figure \ref{fig:zratio_Lambdam32_1_impr}, where the dashed lines are
the results of an analytic calculation which will be reported
elsewhere \cite{baumgartner:2010}. The absence of critical slowing
down for our update algorithm guarantees that reliable results can be
obtained despite the exponentially slow approach to
the continuum.

\section{Summary and outlook}
We have discussed the occurrence of a fermion sign problem in the
context of spontaneous supersymmetry breaking on the lattice and its
relevance for the vanishing of the Witten index, regulated as a path
integral on the lattice. We then argued that with the help of the
fermion loop expansion one can achieve an explicit separation of the
bosonic and fermionic contributions to the path integral in such a way
that the source of the sign problem, namely the cancellation between
the bosonic and fermionic contributions to the partition function with
periodic boundary conditions, is isolated.  The solution of the
fermion sign problem is then achieved by devising an algorithm which
separately samples the bosonic and fermionic contributions to the
partition functions and, in addition, also samples the relative
weights between them, essentially without any critical slowing
down. In such a way one is able to calculate the Witten index on the
lattice without suffering from the fermion sign problem even when the
index vanishes. The absence of critical slowing down is essentially
due to the fact that the algorithm directly samples the massless
Goldstino mode which mediates the tunnelings between the bosonic and
fermionic vacua.

As examples we described in some detail the exact reformulation of the
lattice path integral in terms of fermionic bonds and monomers, and
bosonic bonds for the ${\cal N} = 1$ supersymmetric Wess-Zumino model
in 2 dimensions and ${\cal N} = 2$ supersymmetric quantum
mechanics. For the latter we presented some results from lattice
simulations. For a superpotential with unbroken supersymmetry, we
calculate the energy gap of the lowest bosonic and fermionic
excitation using both the standard discretisation with a fine-tuned
counterterm, as well as a $Q$-exact discretisation which preserves a
linear combination of the two supersymmetries. In the latter case the
boson and fermion spectra are degenerate even at finite lattice
spacing, while in the former case they become degenerate only in the
continuum limit. Finally, we also present lattice calculations of the
Witten index for broken and unbroken supersymmetry using the standard
discretisation with a counter\-term. For both cases we are able to
reproduce the correct Witten index in the continuum limit. For broken
supersymmetry the approach to the continuum limit is exponentially
slow in the lattice spacing. Using the loop formulation and the
fermion worm algorithm \cite{Wenger:2008tq} the exponentially slow
approach as well as the sign problem is no obstacle in practice.

Obviously, the approach presented here is particularly interesting for
the ${\cal N} = 1$ supersymmetric Wess-Zumino model in 2 dimensions,
where so far simulations on the lattice have suffered from the fermion
sign problem \cite{Catterall:2003ae}. Work in this direction is in
progress.

\bibliographystyle{JHEP}
\bibliography{sosmotlwasp}

\end{document}